\documentclass[superscriptaddress,twocolumn,showpacs,preprintnumbers,amsmath,amssymb]{revtex4}

\usepackage{graphicx}
\usepackage{dcolumn} % Align table columns on decimal point
\usepackage{bm} % bold math
\usepackage{color}
\begin{document}

\bibliographystyle{apsrev}

\title{Spilling of electronic states in Pb Quantum Wells}

\date{\today}

\author{M. Ja{\l}ochowski}
\affiliation{Institute of Physics, M. Curie-Sk\l odowska University, 
             Pl. M. Curie-Sk\l odowskiej 1, 20-031 Lublin, Poland}

\author{K. Palot\'as}
\affiliation{Department of Theoretical Physics, Budapest University of
             Technology and Economics, Budafoki \'{u}t 8., H-1111 Budapest, 
             Hungary}

\author{M.~Krawiec} 
  \email{mariusz.krawiec@umcs.pl}
\affiliation{Institute of Physics, M. Curie-Sk\l odowska University, 
             Pl. M. Curie-Sk\l odowskiej 1, 20-031 Lublin, Poland}

\begin{abstract}
Energy-dependent apparent step heights of two-dimensional ultra-thin Pb islands 
grown on the Si(111)6$\times$6Au surface have been investigated by a 
combination of scanning tunneling microscopy, first-principles density
functional theory and the particle in a box model calculations. The apparent 
step height shows the thickness and energy dependent oscillatory behavior, 
which is directly related to the spilling of electron states into the vacuum 
exhibiting a quantum size effect. This has been unambiguously proven by 
extensive first-principles scanning tunneling microscopy and spectroscopy 
simulations. An electronic contribution to the apparent step height is directly 
determined. At certain energies it reaches values as high as a half of the 
atomic contribution. The applicability of the particle in a box model to the 
spilling of electron states is also discussed.
\end{abstract}


\maketitle

%%%%%%%%%%%%%%%%%%%%%%%%%%%%%%%%%%%%%%%%%%%%%%%%%%%%%%%%%%%%%%%%%%%%%%%%%%%%%%%

\section{Introduction}
\label{introduction}

Behavior of metallic quantum wells (QW) on semiconductor surfaces exhibiting 
quantum-size effects (QSE) has been a topic of great fundamental interest for 
many years \cite{Chiang2000,Jia2007,Vazquez2009}. The spatial confinement of 
electrons and the presence of discrete electronic subbands affect electronic 
properties of ultrathin films, as shown for the electrical resistivity 
\cite{Jalochowski1988,Jalochowski1992,Jalochowski1995}, Hall coefficient 
\cite{Jalochowski1996,Vilfan2002}, surface morphology 
\cite{Altfeder1997,Su2001,Otero2002,Jian2003,Unal2007,Binz2008,
Kopciuszynski2015}, surface energy 
\cite{Materzanini2001,Wei2002}, work function \cite{Qi2007,Becker2010}, 
chemical reactivity \cite{Jiang2008}, electron phonon coupling 
\cite{Zhang2005}, superconducting transition temperature 
\cite{Guo2004,Brun2009}, Kondo temperature \cite{Fu2007}, Rashba spin-orbit 
interaction \cite{Slomski2011}, and Friedel oscillations 
\cite{Jia2010,Bouhassoune2014}, to name a few.

Ultrathin Pb layers have been a prominent test laboratory for studying 
electronic quantum effects in nanoscale metallic objects. Theoretical and 
experimental studies indicated that structural and morphological properties of 
ultrathin Pb films are related to QSE. Minimization of the total internal 
energy of Pb island with a thickness dependent QSE electronic component leads 
to the growth of islands with preferential magic thicknesses 
\cite{Budde2000,Yeh2000,Su2001,Hupalo2001,Hupalo2002,Otero2002,Binz2008}. In a 
continuous layer this minimization manifests itself as a variation of the 
measured Pb apparent step height (ASH). Expansion and contraction of the top 
layer were observed in the scattering of He atoms from Pb epitaxial layers on 
Cu(111) \cite{Braun1997} and on Ge(001) \cite{Crottini1997} substrates. The 
density functional theory (DFT) calculations confirmed QSE origin of the 2 ML 
periodicity in the expansion/contraction of Pb layers separation 
\cite{Materzanini2001}. The apparent step height has also been measured by 
x-ray diffraction \cite{Czoschke2003} and scanning tunneling microscopy (STM) 
in a number of QSE systems 
\cite{Su2001,Hupalo2001,Otero2002,Unal2007,Vazquez2009,Calleja2009}. 

It has been proposed that the main reason for oscillation of the ASH is a 
thickness-dependent variation of an electron density outside the quantized 
film. Moreover, it was found that the ASH may depend on the energy, or on the 
STM bias, chosen to probe the sample, since the quantum well state (QWS) 
dominates the electronic structure of the system \cite{Vazquez2009}. In the 
simplest approach, i.e. within a free-electron model applied to a finite 
quantum well, the states "spill out" of the geometrical border of the well 
\cite{Schulte1976}.  The spilling should be larger for the QWSs closer to the 
vacuum level. Thus, it can be expected that the measured thickness of the QW 
depends on the energy of sampling electrons. The STM technique is exceptionally 
well suited to study such effects. In particular, the STM enables the 
exploration of surface electron densities within a broad energy range around 
the Fermi level. 

In the present work we address the issue of bias-dependence of the apparent 
step height in quantum wells, and unambiguously prove that the spilling of 
electron states into the vacuum is governed by energy position of QWS. We 
determine the electronic contribution to the ASH, which is a substantial 
fraction of the atomic contribution, and at certain energies it may reach value
as high as a half of that coming from the atoms arrangement. 
We prove that the 
electronic effects prevail the atomic lattice expansion/contraction, the 
electronic effects may even overwhelm the atomic ones leading to reversing of 
the apparent step height variation upon change of the sample bias in the STM 
experiments.
As an example, we consider Pb quantum wells on 
Si(111)6$\times$6-Au substrate. This substrate was chosen because of the 
specific Pb growth mode achievable \cite{Schmidt2000}. Even the first monatomic 
layer of Pb film has well defined crystalline structure and exhibits clear QSE 
\cite{Jalochowski1992a}. The STM topography measurements reveal the thickness- 
and bias-dependence of the apparent step height. A direct relation between the 
ASH and the spilling of QWSs into the vacuum is unambiguously proven by 
extensive first-principles scanning tunneling microscopy simulations based on 
DFT. Depending on energy the QWS can evanesce in the vacuum at different 
distances. The above observations are explained semi-qualitatively within the 
particle in a box model, which confirms the applicability of this model to QSE 
systems. 
The present study shows that in ultrathin metallic films step 
height determination with STM has to be accompanied by at least simple analysis 
based on finite quantum well model of the QSE states, and their contribution to 
the bias dependent tunneling current.

%%%%%%%%%%%%%%%%%%%%%%%%%%%%%%%%%%%%%%%%%%%%%%%%%%%%%%%%%%%%%%%%%%%%%%%%%%%%%%%

\section{\label{experimental} Experimental and computational details}
 
Samples were prepared in an ultra-high vacuum (UHV) chamber equipped with a STM
(type OMICRON VT) and reflection high energy electron diffraction (RHEED) 
apparatus. The base pressure was $5\times10^{-11}$ mbar. In order to produce 
Si(111)6$\times$6-Au reconstruction about 1.6 ML of Au was deposited onto the 
Si(111)7$\times$7 substrate and annealed for 1 min. at about 950 K, and then 
the temperature was gradually lowered to about 500 K within 3 min. Direct 
resistive heating was used. A series of the Pb films with average thickness 
between 0.5 to 6 ML have been {\it{in situ}} deposited onto the substrate held 
at 170 K in the STM stage. The deposition rate was equal to 0.2 ML of Pb(111) 
per minute. All STM measurements were performed at the sample temperature equal 
to 170 K and the tunneling current equal to 0.1 nA.

Density functional theory calculations were performed within 
Perdew-Burke-Ernzerhof (PBE) \cite{Perdew1996} generalized gradient 
approximation (GGA) using projector-augmented-wave potentials, as implemented 
in VASP (Vienna ab-initio simulation package) \cite{Kresse1996,Kresse1999}. The 
plane wave energy cutoff for all calculations was set to 340 eV, and the
Brillouin zone was sampled by a $5 \times 5 \times 1$ Monkhorst-Pack k-points 
grid \cite{Monkhorst1976}. The spin-orbit interaction was omitted. 

To simplify calculations, the Si(111)$\sqrt{3}\times\sqrt{3}$-Au unit cell has
been adapted, which is locally similar to the Si(111)6$\times$6-Au
\cite{Jalochowski2003}. The Si(111)-Au system has been modeled by 8 Si double 
layers and the reconstruction layer. A vacuum region of 20 \AA\ has been added 
to avoid the interaction between surfaces of the slab. All the atomic positions 
were relaxed, except the bottom layer, until the largest force in any direction 
was below 0.01 eV/\AA. The Si atoms in the bottom layer were fixed at their 
ideal bulk positions and saturated with hydrogen. The lattice constant of Si 
was fixed at the calculated value, 5.47 \AA. 

Based on the obtained electronic structure data of the Si(111)-Au surface 
described above, scanning tunneling microscopy simulations were performed by 
using Bardeen's electron tunneling model \cite{Bardeen1961} implemented in the 
BSKAN code \cite{Hofer2003,Palotas2005,Hofer2005}. The employed tunneling model 
takes into account electronic structures of both the Si(111)-Au surface and 
tungsten tips. We considered blunt and sharp tungsten tip models following Refs.
\cite{Teobaldi2012,Mandi2015} to investigate the effect of tip sharpness on the 
apparent step height. The blunt tip is represented by an adatom adsorbed on the 
hollow site of the W(110) surface, and the sharp tip is modeled as a pyramid of 
three-atoms height on the W(110) surface. More details on the used tip 
geometries and electronic structure calculations can be found in Ref. 
\cite{Teobaldi2012}. Note that the shape of the tip apex structure will 
influence the transmission probability via the Bardeen tunneling matrix 
elements, thus the tunneling current.

%%%%%%%%%%%%%%%%%%%%%%%%%%%%%%%%%%%%%%%%%%%%%%%%%%%%%%%%%%%%%%%%%%%%%%%%%%%%%%%

\section{\label{results} Results and discussion}

\begin{figure}
\includegraphics[width=\linewidth]{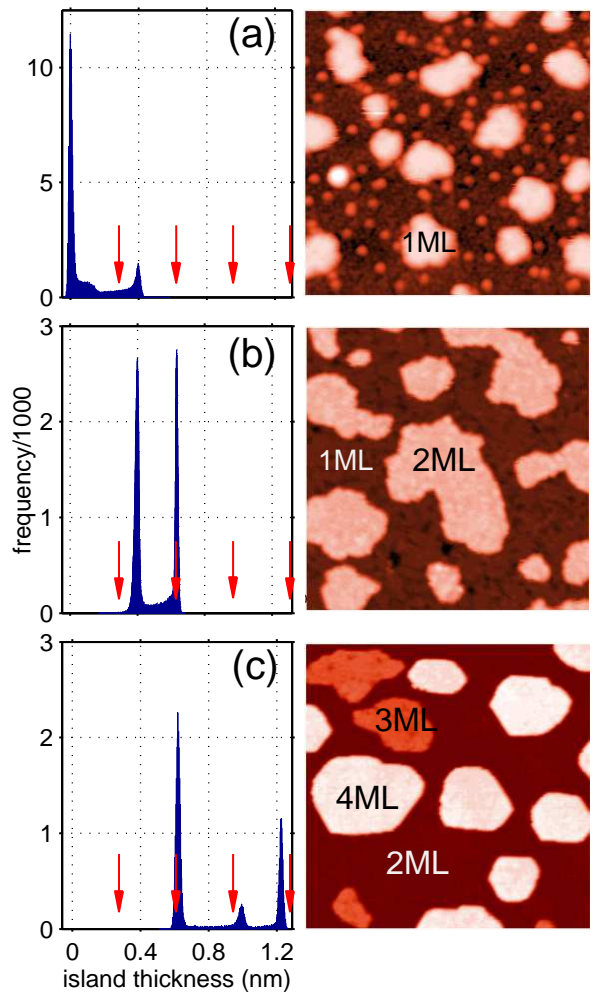}
\caption{\label{fig1} (Color online) Examples of STM topographic images and 
         corresponding histograms of the step height for Pb deposited and 
	 measured at 170 K on Si(111)-(6$\times$6)Au. The Pb coverages are 
	 0.15, 1.4, and 2.8 ML, 
	 their size $30\times30$ $nm^2$, $50\times50$ 
	 $nm^2$, and $100\times100$ $nm^2$,
	 for (a), (b), and (c), respectively. The data were recorded with the 
	 sample bias equal to -2.0 V for the sample (a) and (b), and -2.5 V for 
	 the sample (c). The tunneling current was equal to 0.1 nA. 
	 For all histograms origin of 
	 the x-axis is set at the level of the substrate. Arrows indicate 
	 thicknesses expected for islands with multiples of the of the bulk 
	 Pb(111) ML, equal to 0.285 nm.}
\end{figure}

For coverages studied here Pb formed well resolved 1 and/or 2 ML thick islands. 
Their average lateral size increased with the average film thickness. For 
average coverages less than 1 ML, Fig. \ref{fig1} (a), only 1 ML thick islands 
and single Pb atoms were seen.
After deposition of 1 ML Pb the surface of the film was flat. For average 
coverages between 1 and 2 ML, Fig. \ref{fig1} (b), only 1 ML high islands on 
the 1 ML thick Pb background were seen. No islands were seen after deposition 
of 2 ML. Further increase of the coverage, within the range from 2 to 4 ML, 
resulted in the growth of a mixture of 1 and 2 ML thick islands, as shown in 
Fig. \ref{fig1} (c). At the coverage equal to 4 ML the film was again perfectly 
flat. This process was repeated for the coverages between 4 and 6 ML of Pb. 
Thus, at this particular temperature of deposition, the film grew in a layer by 
layer mode for the coverages up to 2 ML and in the mixed single and double
monolayer modes for the larger coverages. Apparently, the 2, 4, and 6 ML thick 
films are more stable than those with an odd number of monolayers. This finding 
agrees well with the theoretical predictions \cite{Materzanini2001}.

\begin{figure*}[!ht]
\includegraphics[width=0.8\linewidth]{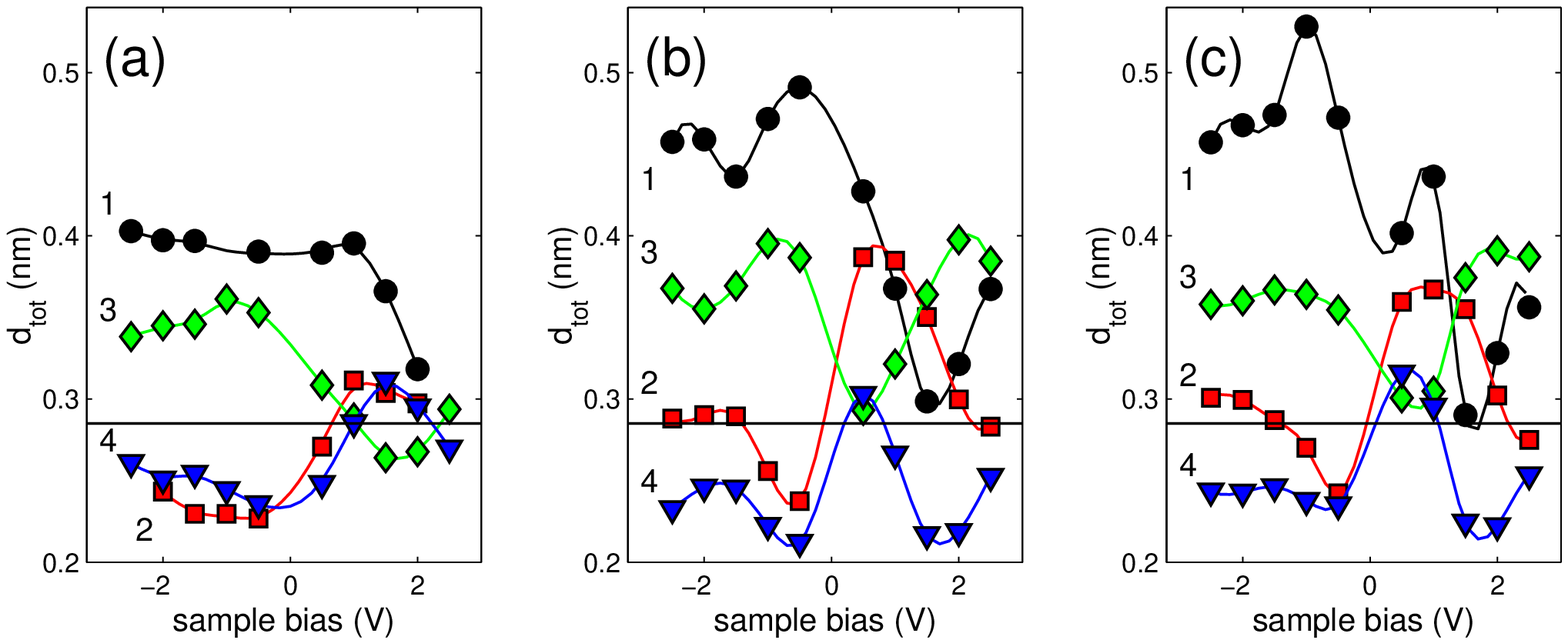}
\caption{ \label{fig2} (Color online) 
	 Apparent step height at various sample bias voltages 
	 determined from the experimental STM topographic images (a) and from 
	 DFT calculations using blunt (b) and sharp (c) tungsten tip models. 
	 The marks $N = 1 - 4$ denote the Pb layer number, i.e. $N$th Pb layer 
	 on a substrate composed of $N-1$ Pb layers. The lines are drawn as eye 
	 guides. The horizontal line denotes the bulk Pb(111) monolayer spacing 
	 $d_0 = 0.285$ nm.}
\end{figure*}

The apparent step height was determined from STM topographic data similar to 
that presented in Fig. \ref{fig1}. A large number of the samples with various 
coverages was produced and scanned with a sample bias V$_{bias}$ ranging from 
-2.5 V to +2.0 V. Next, the inevitable slope of the image background was 
carefully corrected and the image height histograms were calculated. The 
presence of the Si(111) steps (not shown here) allowed the calibration of the 
STM vertical gain. Examples of the histograms are shown in Fig. \ref{fig1}. 
Series of STM scans, similar to that displayed in Fig. \ref{fig1}, produced 
plots of the ASHs as a function of the sample bias. Figure \ref{fig2} (a) shows 
the averaged data of several samples. 
The accuracy of the ASH determination was estimated to be $\pm$ 0.01 nm. Two 
interesting features are seen. First, all curves show strong bias dependence. 
Second, the ASHs on the film with odd number of monolayers behave opposite to 
those on even number of monolayers (including the 1 ML on the substrate). 
For negatively biased samples the ASH of 2 ML and 4 ML Pb, the 
measured apparent step height is much lower than the bulk value $d_0 = 0.285$ 
nm, whereas for 1 ML and 3 ML Pb, it is noticeably larger.
As a consequence, for 
a fixed bias, especially for the negative bias, the ASH varies with 2 ML 
periodicity. 

In order to explain the observed oscillations of the ASH we have performed 
extensive first-principles STM simulations based on DFT. First, the STM 
topography images of the surfaces consisting of a different number of Pb layers 
have been calculated in the constant current mode for various voltages and the 
tunneling current, corresponding to the experimental value. Next, to get the 
ASH of the $N$-th layer, we averaged the topography of the surface with $N$ 
layers over the surface unit cell, and subtracted the corresponding value for 
the surface with $N-1$ layers. 
Such procedure is closely related to the experimentally 
determined ASH, and the results of calculations with the model blunt and sharp 
tips are shown in Fig. \ref{fig2} (b) and (c), respectively. The two tip models 
give similar results for the trends of the ASH, thus we can conclude that the 
bias voltage dependent ASH is governed by the QW states of the surface. This 
also suggests that the possible instability of the tip in STM experiments is 
not expected to qualitatively alter the measured ASH, which might turn out to 
be very useful for future experimental STM studies of QWS.
Clearly, the experimental data of Fig. \ref{fig2} (a) are reproduced 
well, especially for negative sample bias. A poorer agreement in the empty 
state regime is due to known limitations of DFT approach. Note however, that 
the empty-state DFT results coincide with the experimental data, provided they 
are artifically shifted by $\sim + 1$ V (compare panels (a) and 
(b) of Fig. \ref{fig2}). 
Perhaps more sophisticated approaches, like GW or time-dependent
DFT, should give better agreement in the empty state regime.
Slightly higher values calculated for 1 ML Pb, than the experimentally found 
are due to different substrates used in the experiment and in calculations, as 
it was discussed in Sec. \ref{experimental}. Nevertheless, the shape of the 
calculated 1 ML ASH is very similar to the experimental one. 

\begin{figure*}
\includegraphics[width=0.8\linewidth]{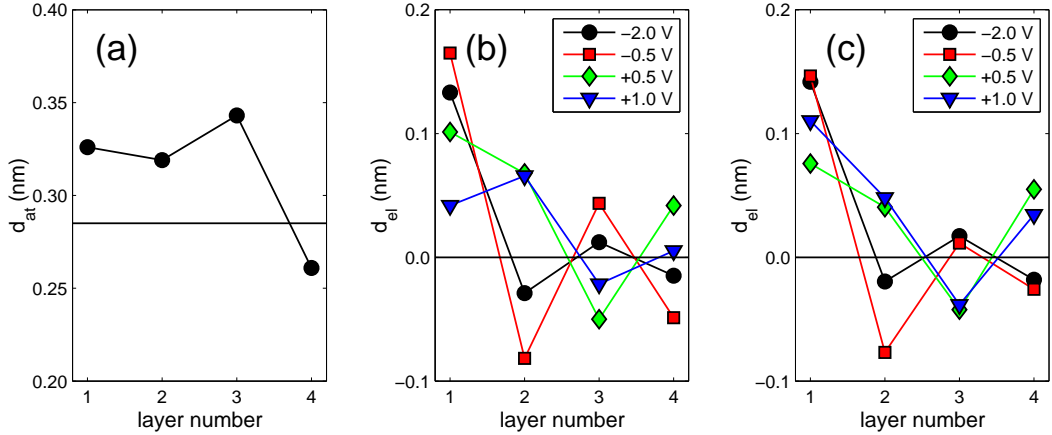}
\caption{\label{fig3} (Color online) Atomic and 
	 electronic contributions to the apparent step height as a function of 
	 the Pb layer number: (a) atomic contribution does not depend on the 
	 tip and on the bias voltage, (b) electronic contribution based on Fig. 
	 \ref{fig2} (b) calculated by the blunt tip, (c) electronic 
	 contribution based on Fig. \ref{fig2} (c) calculated by the sharp tip. 
	 Note that only the electronic part depends on the sample bias. The 
	 horizontal line in (a) denotes the bulk Pb(111) inter-layer spacing.}
\end{figure*}

Strong dependence of the apparent step height on the tunneling bias suggests 
the electronic origin of the observed phenomena. The procedure used in
calculations allows to directly determine both the atomic as well as the 
electronic contributions to ASH. In general, the ASH, $d_{tot}$, can be 
expressed as
\begin{eqnarray}
d_{tot} = d_{at} + d_{el},
\label{eq1}
\end{eqnarray}
where $d_{at}$ is the atomic contribution, simply given by difference of the
z-coordinates of the atoms forming 2 subsequent layers of Pb, and independent 
of the sample bias. $d_{el}$ is the height associated with the electronic 
contribution to the ASH, obviously bias dependent. The atomic and electronic 
contributions to the ASH as a function of the Pb layer number are shown in Fig. 
\ref{fig3}.
The atomic contribution is the same for both considered tip 
models [Fig. \ref{fig3} (a)], whereas the electronic contributions are shown in 
Fig. \ref{fig3} (b) and (c), respectively calculated using the blunt and sharp 
tip models. Similarly to Fig. \ref{fig2}, we find that the two geometrically 
different tip apex structures provide similar trends for the electronic 
contribution of the ASH. Moreover, it is seen that all parts of Fig. \ref{fig3} 
oscillate with periodicity of 2 ML Pb, and the electronic contribution can be 
positive or negative.

Interestingly, the electronic part features higher amplitude of the 
oscillations, than the atomic part. This amplitude depends on the sample bias, 
and increases as the absolute value of sample bias is decreased. As a result, 
at certain bias voltages, the electronic contribution to the ASH may be as 
large as a half of the atomic one, especially for thinner Pb QWs. Compare the 
-0.5 V curve in panel (b) with panel (a). All this confirms that the ASH 
measured by STM is substantially influenced by the electronic effects, i.e. by 
the spilling of the quantum well states into the vacuum. 

To get a deeper and more intuitive insight into the spilling of the QWS and its
impact on the apparent step height, we evoke the one-dimensional model of a 
noninteracting electron gas confined in a finite square QW. Albeit this model 
neglects the layered crystal structure of the film and the presence of the 
substrate, it was previously successfully applied in explanation of the QSE in 
the electron photoemission \cite{Jalochowski1992a}. 
The simple 1D QW model focuses on basic physics related to the 
ASH effects observed in real experiment.
In this model width of the 
quantum well is equal to $N\times{d_{0}}$, with $N$ being the number of the 
Pb monolayers, the effective electron mass $m^{*}$ is equal to $1.002m_{0}$, 
the Fermi energy $E_{F}$ is equal to $9.685$ eV, and the work function is equal 
to $4.35$ eV. For each QW the shape of the wavefunctions $\psi_{n}$ and their 
energy $E_{n}$ were calculated. For explanation of the measured variation of 
the ASH the interesting quantity is the expansion ({\it{spilling}} out) of the 
wavefunctions $\psi_{n}$ outside of the film and its dependence on energy of 
the quantum state.

\begin{figure}
\includegraphics[width=\linewidth]{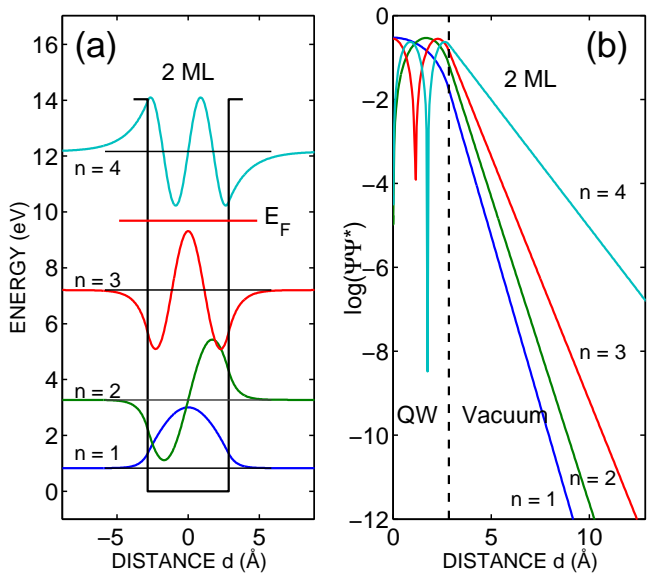}
\caption{\label{fig4} (Color online) The energies of the QW states and the 
         shape of the corresponding normalized wavefunctions $\psi_{n}$ (a). 
	 (b) The logarithm of the one-dimensional probability density 
	 $\log(\psi_{n}\psi_{n}^{*})$ as a function of the distance from the 
	 center of the QW. Notice exponential decay outside the QW. The low 
	 lying states extend much less than the states with the higher energy.}
\end{figure}

\begin{figure}
\includegraphics[width=\linewidth]{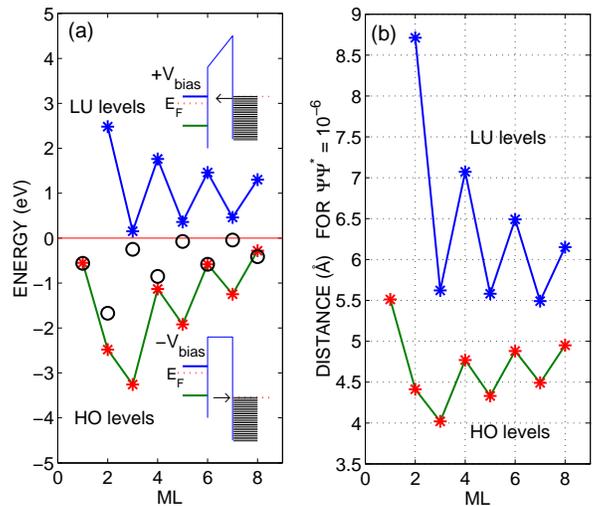}
\caption{\label{fig5} (Color online) The calculated energies (asterisk and plus 
         symbols) of the QW states nearest to the the Fermi energy as a 
	 function of the film thickness (a). Upper and lower insets show 
	 schematic energy diagrams for the 3 ML thick QW during tunneling to 
	 the lowest unoccupied (LU) and highest occupied (HO) states, 
	 respectively. The open circles denote the energies of the QW states 
	 determined in the photoemission experiment \cite{Jalochowski1992a}. 
	 (b) Corresponding distances from the geometric border of the QW at 
	 which the density of the LU and HO states $\psi_{n}\psi_{n}^{*}$ is 
	 equal to $10^{-6}$.}
\end{figure}

Figure \ref{fig4} (a) shows the energies of the QW states and corresponding 
wavefunctions for a QW of the width of 2 ML Pb.
Note the spilling out of the wavefunctions outside the well. As it is displayed 
in Fig. \ref{fig4} (b) in the logarithmic scale, outside the QW the density of 
states of $\psi_{n}$ decays exponentially with the slope of 
$\log(\psi_{n}\psi_{n}^{*})$  depending on the energy of the quantum level. To 
understand the physical origin of the apparent step height oscillation and its 
bias dependence, we recall the Tersoff-Hamann \cite{Tersoff1983} approach to 
the tunneling. 
The Tersoff-Hamann model can be derived from the Bardeen 
tunneling formula by assuming an s-type tip. This popular model, though very 
simple, works reasonably well to understand electron tunneling and STM images 
in a large variety of surface systems with simple electronic structure. To 
provide a simple physical picture for the understanding of the ASH oscillations 
involving the QW states in our 1D QW model, we have used the Tersoff-Hamann 
tunneling model assuming that the transmission probability through the barrier 
and density of states of the tip are constant at all energies. This is 
sufficient to capture the basic physics in QW systems. According to the 
Tersoff-Hamann model, the tunneling current is described by the equation:
\begin{eqnarray}
I(\textbf{\textit{R}})\propto
\sum_{E_{n}>E_{F}}^{E_{n}<E_{F}+eV_{bias}} 
\mid\psi_{n}(\textbf{\textit{R}})\mid^{2}.
\label{eq2}
\end{eqnarray}
Here $\mid\psi_{n}(\textbf{\textit{R}})\mid^{2}$ denotes the local density of 
the state $n$ at the tip position ${\textbf{\textit{R}}}$. In the case of 
extremely thin QW studied here, and the bias ranges applied, only single 
highest occupied (HO) or the lowest unoccupied (LU) QW states  contribute to 
the tunneling current at once. The Eq. \ref{eq2} shows that the tunneling 
current is proportional to the local density of states at the position of STM 
tip, the quantity displayed in the Fig. \ref{fig4} (b). Correspondingly, the 
condition of constant tunneling current to, or from one specific QW level, 
requires adjusting the tip at the distance where $\rho = \psi_{n}\psi_{n}^{*}$ 
is the same. 
Consequently, the tunneling condition was set as typical, with 
experimental sample-tip distance equal 5 \AA\ \cite{Seine1999}. From the figure 
\ref{fig4}(b) follows that such distance, at least for states n = 3 and 4, i.e.
states located within the energy range used in the experiment, corresponds to 
the electron density between $10^{-4}$ and $10^{-8}$. Therefore we have chosen 
the logarithmic mean value of $10^{-6}$ in calculations.
Figure \ref{fig5} shows both, the 
calculated energies of the LU and HO states [panel (a)], and the distances at
which their density is equal to $10^{-6}$ [panel (b)]. 
It is worthwhile to note that experimentally determined positions of the QW 
states \cite{Jalochowski1992a} shown in the Fig. \ref{fig5} (a) as open 
circles, lie close to the theoretical ones and show similar thickness 
dependence. Based on the data of the Fig. \ref{fig5} (b) the origin of the bias 
dependence of the apparent step height shown in Fig. \ref{fig2}, can be 
explained as follows.

\textit{1 ML Pb.} Here only one quantum state is 
accessible for tunneling. According to the data of Fig. \ref{fig5} (a) it lies 
about 0.5 eV below the $E_{F}$ and the tunneling current can flow easily for 
negatively biased sample. 
This state spills far into vacuum and its energy position is
almost independent of $k_{\parallel}$ vector. Moreover, it spreads over a large 
part of the Brillouin zone. For these reasons, the tunneling current is high 
and the STM tip has to be retracted, giving large value of the ASH at negative
sample bias.
For the positively biased sample there is no accessible state and 
in order to keep the same tunneling current, the tip has to be moved closely to 
the sample. As the curve 1 in Fig. \ref{fig2} (a) shows, the measured ASH 
changes from 0.4 nm to about 0.32 nm, for -2 V and +2 V of the sample bias, 
respectively.

\textit{2 ML Pb.} For this QW two quantum states can contribute to 
the tunneling current. Both the HO and the LU states are placed approximately 
symmetrically to the Fermi energy, but their spilling differs considerably, 
compare QW states n = 3 and 4 in Fig. \ref{fig4} (a) and (b).
For n = 4 state (LU) the charge density $10^{-6}$ is obtained
at the distance around 4 \AA\ larger than in the case of n =3 state (HO). 
Consequently, the STM tip has to be retracted for positive sample bias. 
Exactly this behavior is observed in Fig. \ref{fig2} for curves with $N = 2$,
albeit the ASH yields $\sim 1$ \AA.

\textit{3 ML Pb.} 
The behavior of the bias-dependent ASH resembles the situation in
the case of a single Pb layer on the bare substrate ($N = 1$), as the HO state
is located below the Fermi energy. However its energy is almost 3 eV lower than
in the case of $N = 1$. Thus, such low energy should not contribute to the 
current in the bias range used in the experiment. However, this state is more 
dispersive, and the tunneling to the $k_{\parallel} \neq 0$ at lower voltages
dominates. Moreover, the photoemission-determined QW state for 3 ML thick QW 
\cite{Jalochowski1992a} is located just below the Fermi energy. Note, that 
according to the model calculation, this state is located above, albeit very 
close to the E$_F$. In real situation, we can assume that this state will
contribute to current at both sample bias polarizations. Thus the difference in
the ASH for positive and negative sample bias is expected to be smaller, as
comapred to the $N = 1$ case. Indeed this can be seen in Fig. \ref{fig2} (curve
3).

\textit{4 ML Pb.} 
The situation reverses again and resembles the case of $N = 2$.
The HO and LU states are located approximately symmetrically with respect to 
the E$_F$ at energies, which are captured by present STM measurements. 
Moreover, the LU state extends much further into vacuum than the HO, see Fig. 
\ref{fig5} (b), thus the observed ASH is higher at the positive sample bias, as
it is shown in Fig. \ref{fig2} (curve 4).

The above discussion clearly indicates that this simple model explains the most 
important features of the experimental data and of the DFT calculations such as 
variation of the apparent step height with the sample bias and its oscillations 
with 2 ML periodicity. 

%%%%%%%%%%%%%%%%%%%%%%%%%%%%%%%%%%%%%%%%%%%%%%%%%%%%%%%%%%%%%%%%%%%%%%%%%%%%%%%

\section{\label{conclusions} Conclusions}

In conclusion, the apparent step height measured by scanning tunneling
microscopy shows the thickness and energy dependent oscillatory behavior, which 
is directly related to the spilling of quantum well states into the vacuum. The 
electronic contribution to the apparent step height can be as large as a half 
of that coming from the arrangement of atoms. Thus the interpretation of step 
height in ultrathin film determined with scanning tunneling microscopy needs a 
great care and knowledge of its energy dependence. Finally, the simple model of 
the particle in a box contains most of the relevant physics to be successfully 
applied for systems exhibiting quantum size effect.

%%%%%%%%%%%%%%%%%%%%%%%%%%%%%%%%%%%%%%%%%%%%%%%%%%%%%%%%%%%%%%%%%%%%%%%%%%%%%%%

\begin{acknowledgments}
The work has been supported by the National Science Centre (Poland) under Grant 
No. DEC-2014/13/B/ST5/04442. K.P. acknowledges the Hungarian State E\"otv\"os 
Fellowship and the EU COST MP1306 EUSPEC project.
\end{acknowledgments}

%%%%%%%%%%%%%%%%%%%%%%%%%%%%%%%%%%%%%%%%%%%%%%%%%%%%%%%%%%%%%%%%%%%%%%%%%%%%%%%


\begin{thebibliography}{51}
\expandafter\ifx\csname natexlab\endcsname\relax\def\natexlab#1{#1}\fi
\expandafter\ifx\csname bibnamefont\endcsname\relax
  \def\bibnamefont#1{#1}\fi
\expandafter\ifx\csname bibfnamefont\endcsname\relax
  \def\bibfnamefont#1{#1}\fi
\expandafter\ifx\csname citenamefont\endcsname\relax
  \def\citenamefont#1{#1}\fi
\expandafter\ifx\csname url\endcsname\relax
  \def\url#1{\texttt{#1}}\fi
\expandafter\ifx\csname urlprefix\endcsname\relax\def\urlprefix{URL }\fi
\providecommand{\bibinfo}[2]{#2}
\providecommand{\eprint}[2][]{\url{#2}}

\bibitem[{\citenamefont{Chiang}(2000)}]{Chiang2000}
\bibinfo{author}{\bibfnamefont{T.-C.} \bibnamefont{Chiang}},
  \bibinfo{journal}{{S}urf. {S}ci. {R}ep.} \textbf{\bibinfo{volume}{39}},
  \bibinfo{pages}{181} (\bibinfo{year}{2000}).

\bibitem[{\citenamefont{Jia et~al.}(2007)\citenamefont{Jia, Li, Zhang, and
  Xue}}]{Jia2007}
\bibinfo{author}{\bibfnamefont{J.-F.} \bibnamefont{Jia}},
  \bibinfo{author}{\bibfnamefont{S.-C.} \bibnamefont{Li}},
  \bibinfo{author}{\bibfnamefont{Y.-F.} \bibnamefont{Zhang}}, \bibnamefont{and}
  \bibinfo{author}{\bibfnamefont{Q.-K.} \bibnamefont{Xue}},
  \bibinfo{journal}{{J}. {P}hys. {S}oc. {J}pn.} \textbf{\bibinfo{volume}{76}},
  \bibinfo{pages}{2007} (\bibinfo{year}{2007}).

\bibitem[{\citenamefont{{V\'{a}zquez de Parga}
  et~al.}(2009)\citenamefont{{V\'{a}zquez de Parga}, Hinarejos, Calleja,
  Camarero, Otero, and Miranda}}]{Vazquez2009}
\bibinfo{author}{\bibfnamefont{A.~L.} \bibnamefont{{V\'{a}zquez de Parga}}},
  \bibinfo{author}{\bibfnamefont{J.~J.} \bibnamefont{Hinarejos}},
  \bibinfo{author}{\bibfnamefont{F.}~\bibnamefont{Calleja}},
  \bibinfo{author}{\bibfnamefont{J.}~\bibnamefont{Camarero}},
  \bibinfo{author}{\bibfnamefont{R.}~\bibnamefont{Otero}}, \bibnamefont{and}
  \bibinfo{author}{\bibfnamefont{R.}~\bibnamefont{Miranda}},
  \bibinfo{journal}{{S}urf. {S}ci.} \textbf{\bibinfo{volume}{603}},
  \bibinfo{pages}{1389} (\bibinfo{year}{2009}).

\bibitem[{\citenamefont{Ja{\l}ochowski and Bauer}(1988)}]{Jalochowski1988}
\bibinfo{author}{\bibfnamefont{M.}~\bibnamefont{Ja{\l}ochowski}}
  \bibnamefont{and} \bibinfo{author}{\bibfnamefont{E.}~\bibnamefont{Bauer}},
  \bibinfo{journal}{{P}hys. {R}ev. {B}} \textbf{\bibinfo{volume}{38}},
  \bibinfo{pages}{5272} (\bibinfo{year}{1988}).

\bibitem[{\citenamefont{Ja{\l}ochowski
  et~al.}(1992{\natexlab{a}})\citenamefont{Ja{\l}ochowski, Knoppe, Lilienkamp,
  and Bauer}}]{Jalochowski1992}
\bibinfo{author}{\bibfnamefont{M.}~\bibnamefont{Ja{\l}ochowski}},
  \bibinfo{author}{\bibfnamefont{H.}~\bibnamefont{Knoppe}},
  \bibinfo{author}{\bibfnamefont{G.}~\bibnamefont{Lilienkamp}},
  \bibnamefont{and} \bibinfo{author}{\bibfnamefont{E.}~\bibnamefont{Bauer}},
  \bibinfo{journal}{{P}hys. {R}ev. {B}} \textbf{\bibinfo{volume}{46}},
  \bibinfo{pages}{4693} (\bibinfo{year}{1992}{\natexlab{a}}).

\bibitem[{\citenamefont{Ja{\l}ochowski
  et~al.}(1995)\citenamefont{Ja{\l}ochowski, Hoffmann, and
  Bauer}}]{Jalochowski1995}
\bibinfo{author}{\bibfnamefont{M.}~\bibnamefont{Ja{\l}ochowski}},
  \bibinfo{author}{\bibfnamefont{M.}~\bibnamefont{Hoffmann}}, \bibnamefont{and}
  \bibinfo{author}{\bibfnamefont{E.}~\bibnamefont{Bauer}},
  \bibinfo{journal}{{P}hys. {R}ev. {B}} \textbf{\bibinfo{volume}{51}},
  \bibinfo{pages}{7231} (\bibinfo{year}{1995}).

\bibitem[{\citenamefont{Ja{\l}ochowski
  et~al.}(1996)\citenamefont{Ja{\l}ochowski, Hoffmann, and
  Bauer}}]{Jalochowski1996}
\bibinfo{author}{\bibfnamefont{M.}~\bibnamefont{Ja{\l}ochowski}},
  \bibinfo{author}{\bibfnamefont{M.}~\bibnamefont{Hoffmann}}, \bibnamefont{and}
  \bibinfo{author}{\bibfnamefont{E.}~\bibnamefont{Bauer}},
  \bibinfo{journal}{{P}hys. {R}ev. {L}ett.} \textbf{\bibinfo{volume}{76}},
  \bibinfo{pages}{4227} (\bibinfo{year}{1996}).

\bibitem[{\citenamefont{Vilfan et~al.}(2002)\citenamefont{Vilfan, Henzler,
  Pfennigstorf, and Pfnur}}]{Vilfan2002}
\bibinfo{author}{\bibfnamefont{I.}~\bibnamefont{Vilfan}},
  \bibinfo{author}{\bibfnamefont{M.}~\bibnamefont{Henzler}},
  \bibinfo{author}{\bibfnamefont{O.}~\bibnamefont{Pfennigstorf}},
  \bibnamefont{and} \bibinfo{author}{\bibfnamefont{H.}~\bibnamefont{Pfnur}},
  \bibinfo{journal}{{P}hys. {R}ev. {B}} \textbf{\bibinfo{volume}{66}},
  \bibinfo{pages}{241306} (\bibinfo{year}{2002}).

\bibitem[{\citenamefont{Altfeder et~al.}(1997)\citenamefont{Altfeder, Matveev,
  and Chen}}]{Altfeder1997}
\bibinfo{author}{\bibfnamefont{I.}~\bibnamefont{Altfeder}},
  \bibinfo{author}{\bibfnamefont{K.}~\bibnamefont{Matveev}}, \bibnamefont{and}
  \bibinfo{author}{\bibfnamefont{D.}~\bibnamefont{Chen}},
  \bibinfo{journal}{{P}hys. {R}ev. {L}ett.} \textbf{\bibinfo{volume}{78}},
  \bibinfo{pages}{2815} (\bibinfo{year}{1997}).

\bibitem[{\citenamefont{Su et~al.}(2001)\citenamefont{Su, Chang, Jian, Chang,
  Chen, and Tsong}}]{Su2001}
\bibinfo{author}{\bibfnamefont{W.}~\bibnamefont{Su}},
  \bibinfo{author}{\bibfnamefont{S.}~\bibnamefont{Chang}},
  \bibinfo{author}{\bibfnamefont{W.}~\bibnamefont{Jian}},
  \bibinfo{author}{\bibfnamefont{C.}~\bibnamefont{Chang}},
  \bibinfo{author}{\bibfnamefont{L.}~\bibnamefont{Chen}}, \bibnamefont{and}
  \bibinfo{author}{\bibfnamefont{T.}~\bibnamefont{Tsong}},
  \bibinfo{journal}{{P}hys. {R}ev. {L}ett.} \textbf{\bibinfo{volume}{86}},
  \bibinfo{pages}{5116} (\bibinfo{year}{2001}).

\bibitem[{\citenamefont{Otero et~al.}(2002)\citenamefont{Otero, {V\'{a}zquez de
  Parga}, and Miranda}}]{Otero2002}
\bibinfo{author}{\bibfnamefont{R.}~\bibnamefont{Otero}},
  \bibinfo{author}{\bibfnamefont{A.~L.} \bibnamefont{{V\'{a}zquez de Parga}}},
  \bibnamefont{and} \bibinfo{author}{\bibfnamefont{R.}~\bibnamefont{Miranda}},
  \bibinfo{journal}{{P}hys. {R}ev. {B}} \textbf{\bibinfo{volume}{66}},
  \bibinfo{pages}{115401} (\bibinfo{year}{2002}).

\bibitem[{\citenamefont{Jian et~al.}(2003)\citenamefont{Jian, Su, Chang, and
  Tsong}}]{Jian2003}
\bibinfo{author}{\bibfnamefont{W.}~\bibnamefont{Jian}},
  \bibinfo{author}{\bibfnamefont{W.}~\bibnamefont{Su}},
  \bibinfo{author}{\bibfnamefont{C.}~\bibnamefont{Chang}}, \bibnamefont{and}
  \bibinfo{author}{\bibfnamefont{T.}~\bibnamefont{Tsong}},
  \bibinfo{journal}{{P}hys. {R}ev. {L}ett.} \textbf{\bibinfo{volume}{90}},
  \bibinfo{pages}{196603} (\bibinfo{year}{2003}).

\bibitem[{\citenamefont{Unal et~al.}(2007)\citenamefont{Unal, Gin, Han, Liu,
  Jing, Layson, Jenks, Evans, and Thiel}}]{Unal2007}
\bibinfo{author}{\bibfnamefont{B.}~\bibnamefont{Unal}},
  \bibinfo{author}{\bibfnamefont{F.}~\bibnamefont{Gin}},
  \bibinfo{author}{\bibfnamefont{Y.}~\bibnamefont{Han}},
  \bibinfo{author}{\bibfnamefont{D.-J.} \bibnamefont{Liu}},
  \bibinfo{author}{\bibfnamefont{D.}~\bibnamefont{Jing}},
  \bibinfo{author}{\bibfnamefont{A.~R.} \bibnamefont{Layson}},
  \bibinfo{author}{\bibfnamefont{C.~J.} \bibnamefont{Jenks}},
  \bibinfo{author}{\bibfnamefont{J.~W.} \bibnamefont{Evans}}, \bibnamefont{and}
  \bibinfo{author}{\bibfnamefont{P.~A.} \bibnamefont{Thiel}},
  \bibinfo{journal}{{P}hys. {R}ev. {B}} \textbf{\bibinfo{volume}{76}},
  \bibinfo{pages}{195410} (\bibinfo{year}{2007}).

\bibitem[{\citenamefont{Binz et~al.}(2008)\citenamefont{Binz, Hupalo, and
  Tringides}}]{Binz2008}
\bibinfo{author}{\bibfnamefont{S.~M.} \bibnamefont{Binz}},
  \bibinfo{author}{\bibfnamefont{M.}~\bibnamefont{Hupalo}}, \bibnamefont{and}
  \bibinfo{author}{\bibfnamefont{M.~C.} \bibnamefont{Tringides}},
  \bibinfo{journal}{{P}hys. {R}ev. {B}} \textbf{\bibinfo{volume}{78}},
  \bibinfo{pages}{193407} (\bibinfo{year}{2008}).

\bibitem[{\citenamefont{Kopciuszy\'{n}ski
  et~al.}(2015)\citenamefont{Kopciuszy\'{n}ski, Dyniec, Krawiec,
  Ja{\l}ochowski, and Zdyb}}]{Kopciuszynski2015}
\bibinfo{author}{\bibfnamefont{M.}~\bibnamefont{Kopciuszy\'{n}ski}},
  \bibinfo{author}{\bibfnamefont{P.}~\bibnamefont{Dyniec}},
  \bibinfo{author}{\bibfnamefont{M.}~\bibnamefont{Krawiec}},
  \bibinfo{author}{\bibfnamefont{M.}~\bibnamefont{Ja{\l}ochowski}},
  \bibnamefont{and} \bibinfo{author}{\bibfnamefont{R.}~\bibnamefont{Zdyb}},
  \bibinfo{journal}{{A}ppl. {S}urf. {S}ci.} \textbf{\bibinfo{volume}{331}},
  \bibinfo{pages}{512} (\bibinfo{year}{2015}).

\bibitem[{\citenamefont{Materzanini et~al.}(2001)\citenamefont{Materzanini,
  Saalfrank, and Lindan}}]{Materzanini2001}
\bibinfo{author}{\bibfnamefont{G.}~\bibnamefont{Materzanini}},
  \bibinfo{author}{\bibfnamefont{P.}~\bibnamefont{Saalfrank}},
  \bibnamefont{and} \bibinfo{author}{\bibfnamefont{P.}~\bibnamefont{Lindan}},
  \bibinfo{journal}{{P}hys. {R}ev. {B}} \textbf{\bibinfo{volume}{63}},
  \bibinfo{pages}{235405} (\bibinfo{year}{2001}).

\bibitem[{\citenamefont{Wei and Chou}(2002)}]{Wei2002}
\bibinfo{author}{\bibfnamefont{C.}~\bibnamefont{Wei}} \bibnamefont{and}
  \bibinfo{author}{\bibfnamefont{M.}~\bibnamefont{Chou}},
  \bibinfo{journal}{{P}hys. {R}ev. {B}} \textbf{\bibinfo{volume}{66}},
  \bibinfo{pages}{233408} (\bibinfo{year}{2002}).

\bibitem[{\citenamefont{Qi et~al.}(2007)\citenamefont{Qi, Ma, Jiang, Ji, Fu,
  Jia, Xue, and Zhang}}]{Qi2007}
\bibinfo{author}{\bibfnamefont{Y.}~\bibnamefont{Qi}},
  \bibinfo{author}{\bibfnamefont{X.}~\bibnamefont{Ma}},
  \bibinfo{author}{\bibfnamefont{P.}~\bibnamefont{Jiang}},
  \bibinfo{author}{\bibfnamefont{S.}~\bibnamefont{Ji}},
  \bibinfo{author}{\bibfnamefont{Y.}~\bibnamefont{Fu}},
  \bibinfo{author}{\bibfnamefont{J.}~\bibnamefont{Jia}},
  \bibinfo{author}{\bibfnamefont{Q.}~\bibnamefont{Xue}}, \bibnamefont{and}
  \bibinfo{author}{\bibfnamefont{S.}~\bibnamefont{Zhang}},
  \bibinfo{journal}{{A}ppl. {P}hys. {L}ett.} \textbf{\bibinfo{volume}{90}},
  \bibinfo{pages}{013109} (\bibinfo{year}{2007}).

\bibitem[{\citenamefont{Becker and Berndt}(2010)}]{Becker2010}
\bibinfo{author}{\bibfnamefont{M.}~\bibnamefont{Becker}} \bibnamefont{and}
  \bibinfo{author}{\bibfnamefont{R.}~\bibnamefont{Berndt}},
  \bibinfo{journal}{{A}ppl. {P}hys. {L}ett.} \textbf{\bibinfo{volume}{96}},
  \bibinfo{pages}{033112} (\bibinfo{year}{2010}).

\bibitem[{\citenamefont{Jiang et~al.}(2008)\citenamefont{Jiang, Ma, Ning, Song,
  Chen, Jia, and Xue}}]{Jiang2008}
\bibinfo{author}{\bibfnamefont{P.}~\bibnamefont{Jiang}},
  \bibinfo{author}{\bibfnamefont{X.}~\bibnamefont{Ma}},
  \bibinfo{author}{\bibfnamefont{Y.}~\bibnamefont{Ning}},
  \bibinfo{author}{\bibfnamefont{C.}~\bibnamefont{Song}},
  \bibinfo{author}{\bibfnamefont{X.}~\bibnamefont{Chen}},
  \bibinfo{author}{\bibfnamefont{J.-F.} \bibnamefont{Jia}}, \bibnamefont{and}
  \bibinfo{author}{\bibfnamefont{Q.-K.} \bibnamefont{Xue}},
  \bibinfo{journal}{{J}. {A}m. {C}hem. {S}oc.} \textbf{\bibinfo{volume}{130}},
  \bibinfo{pages}{7790} (\bibinfo{year}{2008}).

\bibitem[{\citenamefont{Zhang et~al.}(2005)\citenamefont{Zhang, Jia, Han, Tang,
  Shen, Guo, Qiu, and Xue}}]{Zhang2005}
\bibinfo{author}{\bibfnamefont{Y.}~\bibnamefont{Zhang}},
  \bibinfo{author}{\bibfnamefont{J.}~\bibnamefont{Jia}},
  \bibinfo{author}{\bibfnamefont{T.}~\bibnamefont{Han}},
  \bibinfo{author}{\bibfnamefont{Z.}~\bibnamefont{Tang}},
  \bibinfo{author}{\bibfnamefont{Q.}~\bibnamefont{Shen}},
  \bibinfo{author}{\bibfnamefont{Y.}~\bibnamefont{Guo}},
  \bibinfo{author}{\bibfnamefont{Z.}~\bibnamefont{Qiu}}, \bibnamefont{and}
  \bibinfo{author}{\bibfnamefont{Q.}~\bibnamefont{Xue}},
  \bibinfo{journal}{{P}hys. {R}ev. {L}ett.} \textbf{\bibinfo{volume}{95}},
  \bibinfo{pages}{096802} (\bibinfo{year}{2005}).

\bibitem[{\citenamefont{Guo et~al.}(2004)\citenamefont{Guo, Zhang, Bao, Han,
  Tang, Zhang, Zhu, Wang, Niu, Qiu et~al.}}]{Guo2004}
\bibinfo{author}{\bibfnamefont{Y.}~\bibnamefont{Guo}},
  \bibinfo{author}{\bibfnamefont{Y.-F.} \bibnamefont{Zhang}},
  \bibinfo{author}{\bibfnamefont{X.-Y.} \bibnamefont{Bao}},
  \bibinfo{author}{\bibfnamefont{T.-Z.} \bibnamefont{Han}},
  \bibinfo{author}{\bibfnamefont{Z.}~\bibnamefont{Tang}},
  \bibinfo{author}{\bibfnamefont{Z.-X.} \bibnamefont{Zhang}},
  \bibinfo{author}{\bibfnamefont{W.-G.} \bibnamefont{Zhu}},
  \bibinfo{author}{\bibfnamefont{E.}~\bibnamefont{Wang}},
  \bibinfo{author}{\bibfnamefont{Q.}~\bibnamefont{Niu}},
  \bibinfo{author}{\bibfnamefont{Z.}~\bibnamefont{Qiu}}, \bibnamefont{et~al.},
  \bibinfo{journal}{{S}cience} \textbf{\bibinfo{volume}{306}},
  \bibinfo{pages}{1915} (\bibinfo{year}{2004}).

\bibitem[{\citenamefont{Brun et~al.}(2009)\citenamefont{Brun, Hong, Pattney,
  Sklyadneva, Heid, Echenique, Bohnen, Chulkov, and Schneider}}]{Brun2009}
\bibinfo{author}{\bibfnamefont{C.}~\bibnamefont{Brun}},
  \bibinfo{author}{\bibfnamefont{I.}~\bibnamefont{Hong}},
  \bibinfo{author}{\bibfnamefont{F.}~\bibnamefont{Pattney}},
  \bibinfo{author}{\bibfnamefont{I.}~\bibnamefont{Sklyadneva}},
  \bibinfo{author}{\bibfnamefont{R.}~\bibnamefont{Heid}},
  \bibinfo{author}{\bibfnamefont{P.}~\bibnamefont{Echenique}},
  \bibinfo{author}{\bibfnamefont{K.}~\bibnamefont{Bohnen}},
  \bibinfo{author}{\bibfnamefont{E.}~\bibnamefont{Chulkov}}, \bibnamefont{and}
  \bibinfo{author}{\bibfnamefont{W.}~\bibnamefont{Schneider}},
  \bibinfo{journal}{{P}hys. {R}ev. {L}ett.} \textbf{\bibinfo{volume}{102}},
  \bibinfo{pages}{207002} (\bibinfo{year}{2009}).

\bibitem[{\citenamefont{Fu et~al.}(2007)\citenamefont{Fu, Ji, Chen, Ma, Wu,
  Wang, Duan, Qiu, Sun, Zhang et~al.}}]{Fu2007}
\bibinfo{author}{\bibfnamefont{Y.-S.} \bibnamefont{Fu}},
  \bibinfo{author}{\bibfnamefont{S.-H.} \bibnamefont{Ji}},
  \bibinfo{author}{\bibfnamefont{X.}~\bibnamefont{Chen}},
  \bibinfo{author}{\bibfnamefont{X.-C.} \bibnamefont{Ma}},
  \bibinfo{author}{\bibfnamefont{R.}~\bibnamefont{Wu}},
  \bibinfo{author}{\bibfnamefont{C.-C.} \bibnamefont{Wang}},
  \bibinfo{author}{\bibfnamefont{W.-H.} \bibnamefont{Duan}},
  \bibinfo{author}{\bibfnamefont{X.-H.} \bibnamefont{Qiu}},
  \bibinfo{author}{\bibfnamefont{B.}~\bibnamefont{Sun}},
  \bibinfo{author}{\bibfnamefont{P.}~\bibnamefont{Zhang}},
  \bibnamefont{et~al.}, \bibinfo{journal}{{P}hys. {R}ev. {L}ett.}
  \textbf{\bibinfo{volume}{99}}, \bibinfo{pages}{256601}
  (\bibinfo{year}{2007}).

\bibitem[{\citenamefont{Slomski et~al.}(2011)\citenamefont{Slomski, Landolt,
  Meier, Patthey, Bihlmayer, Osterwalder, and Dil}}]{Slomski2011}
\bibinfo{author}{\bibfnamefont{B.}~\bibnamefont{Slomski}},
  \bibinfo{author}{\bibfnamefont{G.}~\bibnamefont{Landolt}},
  \bibinfo{author}{\bibfnamefont{F.}~\bibnamefont{Meier}},
  \bibinfo{author}{\bibfnamefont{L.}~\bibnamefont{Patthey}},
  \bibinfo{author}{\bibfnamefont{G.}~\bibnamefont{Bihlmayer}},
  \bibinfo{author}{\bibfnamefont{J.}~\bibnamefont{Osterwalder}},
  \bibnamefont{and} \bibinfo{author}{\bibfnamefont{J.}~\bibnamefont{Dil}},
  \bibinfo{journal}{{P}hys. {R}ev. {B}} \textbf{\bibinfo{volume}{84}},
  \bibinfo{pages}{193406} (\bibinfo{year}{2011}).

\bibitem[{\citenamefont{Jia et~al.}(2010)\citenamefont{Jia, Wu, Li, Einstein,
  Weitering, and Zhang}}]{Jia2010}
\bibinfo{author}{\bibfnamefont{Y.}~\bibnamefont{Jia}},
  \bibinfo{author}{\bibfnamefont{B.}~\bibnamefont{Wu}},
  \bibinfo{author}{\bibfnamefont{C.}~\bibnamefont{Li}},
  \bibinfo{author}{\bibfnamefont{T.}~\bibnamefont{Einstein}},
  \bibinfo{author}{\bibfnamefont{H.}~\bibnamefont{Weitering}},
  \bibnamefont{and} \bibinfo{author}{\bibfnamefont{Z.}~\bibnamefont{Zhang}},
  \bibinfo{journal}{{P}hys. {R}ev. {L}ett.} \textbf{\bibinfo{volume}{105}},
  \bibinfo{pages}{066101} (\bibinfo{year}{2010}).

\bibitem[{\citenamefont{Bouhassoune et~al.}(2014)\citenamefont{Bouhassoune,
  Zimmermann, Mavropoulos, Wortmann, Dederichs, Bl\"{u}gel, and
  Lounis}}]{Bouhassoune2014}
\bibinfo{author}{\bibfnamefont{M.}~\bibnamefont{Bouhassoune}},
  \bibinfo{author}{\bibfnamefont{B.}~\bibnamefont{Zimmermann}},
  \bibinfo{author}{\bibfnamefont{P.}~\bibnamefont{Mavropoulos}},
  \bibinfo{author}{\bibfnamefont{D.}~\bibnamefont{Wortmann}},
  \bibinfo{author}{\bibfnamefont{P.}~\bibnamefont{Dederichs}},
  \bibinfo{author}{\bibfnamefont{S.}~\bibnamefont{Bl\"{u}gel}},
  \bibnamefont{and} \bibinfo{author}{\bibfnamefont{S.}~\bibnamefont{Lounis}},
  \bibinfo{journal}{{N}ature {C}ommun.} \textbf{\bibinfo{volume}{5}},
  \bibinfo{pages}{5558} (\bibinfo{year}{2014}).

\bibitem[{\citenamefont{Budde et~al.}(2000)\citenamefont{Budde, Abram, Yeh, and
  Tringides}}]{Budde2000}
\bibinfo{author}{\bibfnamefont{K.}~\bibnamefont{Budde}},
  \bibinfo{author}{\bibfnamefont{E.}~\bibnamefont{Abram}},
  \bibinfo{author}{\bibfnamefont{V.}~\bibnamefont{Yeh}}, \bibnamefont{and}
  \bibinfo{author}{\bibfnamefont{M.}~\bibnamefont{Tringides}},
  \bibinfo{journal}{{P}hys. {R}ev. {B}} \textbf{\bibinfo{volume}{61}},
  \bibinfo{pages}{R10602} (\bibinfo{year}{2000}).

\bibitem[{\citenamefont{Yeh et~al.}(2000)\citenamefont{Yeh, Berbil-Bautista,
  Wang, Ho, and Tringides}}]{Yeh2000}
\bibinfo{author}{\bibfnamefont{V.}~\bibnamefont{Yeh}},
  \bibinfo{author}{\bibfnamefont{L.}~\bibnamefont{Berbil-Bautista}},
  \bibinfo{author}{\bibfnamefont{C.}~\bibnamefont{Wang}},
  \bibinfo{author}{\bibfnamefont{K.}~\bibnamefont{Ho}}, \bibnamefont{and}
  \bibinfo{author}{\bibfnamefont{M.}~\bibnamefont{Tringides}},
  \bibinfo{journal}{{P}hys. {R}ev. {L}ett.} \textbf{\bibinfo{volume}{85}},
  \bibinfo{pages}{5158} (\bibinfo{year}{2000}).

\bibitem[{\citenamefont{Hupalo et~al.}(2001)\citenamefont{Hupalo, Yeh,
  Berbil-Bautista, Kremer, Abram, and Tringides}}]{Hupalo2001}
\bibinfo{author}{\bibfnamefont{M.}~\bibnamefont{Hupalo}},
  \bibinfo{author}{\bibfnamefont{V.}~\bibnamefont{Yeh}},
  \bibinfo{author}{\bibfnamefont{L.}~\bibnamefont{Berbil-Bautista}},
  \bibinfo{author}{\bibfnamefont{S.}~\bibnamefont{Kremer}},
  \bibinfo{author}{\bibfnamefont{E.}~\bibnamefont{Abram}}, \bibnamefont{and}
  \bibinfo{author}{\bibfnamefont{M.}~\bibnamefont{Tringides}},
  \bibinfo{journal}{{P}hys. {R}ev. {B}} \textbf{\bibinfo{volume}{64}},
  \bibinfo{pages}{155307} (\bibinfo{year}{2001}).

\bibitem[{\citenamefont{Hupalo and Tringides}(2002)}]{Hupalo2002}
\bibinfo{author}{\bibfnamefont{M.}~\bibnamefont{Hupalo}} \bibnamefont{and}
  \bibinfo{author}{\bibfnamefont{M.}~\bibnamefont{Tringides}},
  \bibinfo{journal}{{P}hys. {R}ev. {B}} \textbf{\bibinfo{volume}{65}},
  \bibinfo{pages}{115406} (\bibinfo{year}{2002}).

\bibitem[{\citenamefont{Braun and Toennies}(1997)}]{Braun1997}
\bibinfo{author}{\bibfnamefont{J.}~\bibnamefont{Braun}} \bibnamefont{and}
  \bibinfo{author}{\bibfnamefont{J.}~\bibnamefont{Toennies}},
  \bibinfo{journal}{{S}urf. {S}ci.} \textbf{\bibinfo{volume}{384}},
  \bibinfo{pages}{L858} (\bibinfo{year}{1997}).

\bibitem[{\citenamefont{Crottini et~al.}(1997)\citenamefont{Crottini, Cvetko,
  Floreano, Gotter, Morgante, and Tommasini}}]{Crottini1997}
\bibinfo{author}{\bibfnamefont{A.}~\bibnamefont{Crottini}},
  \bibinfo{author}{\bibfnamefont{D.}~\bibnamefont{Cvetko}},
  \bibinfo{author}{\bibfnamefont{L.}~\bibnamefont{Floreano}},
  \bibinfo{author}{\bibfnamefont{R.}~\bibnamefont{Gotter}},
  \bibinfo{author}{\bibfnamefont{A.}~\bibnamefont{Morgante}}, \bibnamefont{and}
  \bibinfo{author}{\bibfnamefont{F.}~\bibnamefont{Tommasini}},
  \bibinfo{journal}{{P}hys. {R}ev. {L}ett.} \textbf{\bibinfo{volume}{79}},
  \bibinfo{pages}{1527} (\bibinfo{year}{1997}).

\bibitem[{\citenamefont{Czoschke et~al.}(2003)\citenamefont{Czoschke, Hong,
  Basile, and Chiang}}]{Czoschke2003}
\bibinfo{author}{\bibfnamefont{P.}~\bibnamefont{Czoschke}},
  \bibinfo{author}{\bibfnamefont{H.}~\bibnamefont{Hong}},
  \bibinfo{author}{\bibfnamefont{L.}~\bibnamefont{Basile}}, \bibnamefont{and}
  \bibinfo{author}{\bibfnamefont{T.-C.} \bibnamefont{Chiang}},
  \bibinfo{journal}{{P}hys. {R}ev. {L}ett.} \textbf{\bibinfo{volume}{91}},
  \bibinfo{pages}{226801} (\bibinfo{year}{2003}).

\bibitem[{\citenamefont{Calleja et~al.}(2009)\citenamefont{Calleja,
  {V\'{a}zquez de Parga}, Anglada, Hinarejos, Miranda, and
  Yndurain}}]{Calleja2009}
\bibinfo{author}{\bibfnamefont{F.}~\bibnamefont{Calleja}},
  \bibinfo{author}{\bibfnamefont{A.}~\bibnamefont{{V\'{a}zquez de Parga}}},
  \bibinfo{author}{\bibfnamefont{E.}~\bibnamefont{Anglada}},
  \bibinfo{author}{\bibfnamefont{J.}~\bibnamefont{Hinarejos}},
  \bibinfo{author}{\bibfnamefont{R.}~\bibnamefont{Miranda}}, \bibnamefont{and}
  \bibinfo{author}{\bibfnamefont{F.}~\bibnamefont{Yndurain}},
  \bibinfo{journal}{{N}ew {J}. {P}hys.} \textbf{\bibinfo{volume}{11}},
  \bibinfo{pages}{123003} (\bibinfo{year}{2009}).

\bibitem[{\citenamefont{Schulte}(1976)}]{Schulte1976}
\bibinfo{author}{\bibfnamefont{F.}~\bibnamefont{Schulte}},
  \bibinfo{journal}{{S}urf. {S}ci.} \textbf{\bibinfo{volume}{55}},
  \bibinfo{pages}{427} (\bibinfo{year}{1976}).

\bibitem[{\citenamefont{Schmidt and Bauer}(2000)}]{Schmidt2000}
\bibinfo{author}{\bibfnamefont{T.}~\bibnamefont{Schmidt}} \bibnamefont{and}
  \bibinfo{author}{\bibfnamefont{E.}~\bibnamefont{Bauer}},
  \bibinfo{journal}{{P}hys. {R}ev. {B}} \textbf{\bibinfo{volume}{62}},
  \bibinfo{pages}{15815} (\bibinfo{year}{2000}).

\bibitem[{\citenamefont{Ja{\l}ochowski
  et~al.}(1992{\natexlab{b}})\citenamefont{Ja{\l}ochowski, Bauer, Knoppe, and
  Lilienkamp}}]{Jalochowski1992a}
\bibinfo{author}{\bibfnamefont{M.}~\bibnamefont{Ja{\l}ochowski}},
  \bibinfo{author}{\bibfnamefont{E.}~\bibnamefont{Bauer}},
  \bibinfo{author}{\bibfnamefont{H.}~\bibnamefont{Knoppe}}, \bibnamefont{and}
  \bibinfo{author}{\bibfnamefont{G.}~\bibnamefont{Lilienkamp}},
  \bibinfo{journal}{{P}hys. {R}ev. {B}} \textbf{\bibinfo{volume}{45}},
  \bibinfo{pages}{13607} (\bibinfo{year}{1992}{\natexlab{b}}).

\bibitem[{\citenamefont{Perdew et~al.}(1996)\citenamefont{Perdew, Burke, and
  Ernzerhof}}]{Perdew1996}
\bibinfo{author}{\bibfnamefont{J.~P.} \bibnamefont{Perdew}},
  \bibinfo{author}{\bibfnamefont{K.}~\bibnamefont{Burke}}, \bibnamefont{and}
  \bibinfo{author}{\bibfnamefont{M.}~\bibnamefont{Ernzerhof}},
  \bibinfo{journal}{{P}hys. {R}ev. {L}ett.} \textbf{\bibinfo{volume}{77}},
  \bibinfo{pages}{2865} (\bibinfo{year}{1996}).

\bibitem[{\citenamefont{Kresse and Furthm\"{u}ller}(1996)}]{Kresse1996}
\bibinfo{author}{\bibfnamefont{G.}~\bibnamefont{Kresse}} \bibnamefont{and}
  \bibinfo{author}{\bibfnamefont{J.}~\bibnamefont{Furthm\"{u}ller}},
  \bibinfo{journal}{{P}hys. {R}ev. {B}} \textbf{\bibinfo{volume}{54}},
  \bibinfo{pages}{11169} (\bibinfo{year}{1996}).

\bibitem[{\citenamefont{Kresse and Joubert}(1999)}]{Kresse1999}
\bibinfo{author}{\bibfnamefont{G.}~\bibnamefont{Kresse}} \bibnamefont{and}
  \bibinfo{author}{\bibfnamefont{D.}~\bibnamefont{Joubert}},
  \bibinfo{journal}{{P}hys. {R}ev. {B}} \textbf{\bibinfo{volume}{59}},
  \bibinfo{pages}{1758} (\bibinfo{year}{1999}).

\bibitem[{\citenamefont{Monkhorst and Pack}(1976)}]{Monkhorst1976}
\bibinfo{author}{\bibfnamefont{H.~J.} \bibnamefont{Monkhorst}}
  \bibnamefont{and} \bibinfo{author}{\bibfnamefont{J.~D.} \bibnamefont{Pack}},
  \bibinfo{journal}{{P}hys. {R}ev. {B}} \textbf{\bibinfo{volume}{13}},
  \bibinfo{pages}{5188} (\bibinfo{year}{1976}).

\bibitem[{\citenamefont{Ja{\l}ochowski}(2003)}]{Jalochowski2003}
\bibinfo{author}{\bibfnamefont{M.}~\bibnamefont{Ja{\l}ochowski}},
  \bibinfo{journal}{{P}rog. {S}urf. {S}ci.} \textbf{\bibinfo{volume}{74}},
  \bibinfo{pages}{97} (\bibinfo{year}{2003}).

\bibitem[{\citenamefont{Bardeen}(1961)}]{Bardeen1961}
\bibinfo{author}{\bibfnamefont{J.}~\bibnamefont{Bardeen}},
  \bibinfo{journal}{{P}hys. {R}ev. {L}ett.} \textbf{\bibinfo{volume}{6}},
  \bibinfo{pages}{57} (\bibinfo{year}{1961}).

\bibitem[{\citenamefont{Hofer}(2003)}]{Hofer2003}
\bibinfo{author}{\bibfnamefont{W.~A.} \bibnamefont{Hofer}},
  \bibinfo{journal}{{P}rog. {S}urf. {S}ci.} \textbf{\bibinfo{volume}{71}},
  \bibinfo{pages}{147} (\bibinfo{year}{2003}).

\bibitem[{\citenamefont{Palot\'{a}s and Hofer}(2005)}]{Palotas2005}
\bibinfo{author}{\bibfnamefont{K.}~\bibnamefont{Palot\'{a}s}} \bibnamefont{and}
  \bibinfo{author}{\bibfnamefont{W.~A.} \bibnamefont{Hofer}},
  \bibinfo{journal}{{J}. {P}hys.: {C}ondens. {M}atter}
  \textbf{\bibinfo{volume}{17}}, \bibinfo{pages}{2705} (\bibinfo{year}{2005}).

\bibitem[{\citenamefont{Hofer and Garcia-Lekue}(2005)}]{Hofer2005}
\bibinfo{author}{\bibfnamefont{W.~A.} \bibnamefont{Hofer}} \bibnamefont{and}
  \bibinfo{author}{\bibfnamefont{A.}~\bibnamefont{Garcia-Lekue}},
  \bibinfo{journal}{{P}hys. {R}ev. {B}} \textbf{\bibinfo{volume}{71}},
  \bibinfo{pages}{085401} (\bibinfo{year}{2005}).

\bibitem[{\citenamefont{Teobaldi et~al.}(2012)\citenamefont{Teobaldi, Inami,
  Kanasaki, Tanimura, and Shluger}}]{Teobaldi2012}
\bibinfo{author}{\bibfnamefont{G.}~\bibnamefont{Teobaldi}},
  \bibinfo{author}{\bibfnamefont{E.}~\bibnamefont{Inami}},
  \bibinfo{author}{\bibfnamefont{J.}~\bibnamefont{Kanasaki}},
  \bibinfo{author}{\bibfnamefont{K.}~\bibnamefont{Tanimura}}, \bibnamefont{and}
  \bibinfo{author}{\bibfnamefont{A.~L.} \bibnamefont{Shluger}},
  \bibinfo{journal}{{P}hys. {R}ev. {B}} \textbf{\bibinfo{volume}{85}},
  \bibinfo{pages}{085433} (\bibinfo{year}{2012}).

\bibitem[{\citenamefont{M\'{a}ndi et~al.}(2015)\citenamefont{M\'{a}ndi,
  Teobaldi, and Palot\'{a}s}}]{Mandi2015}
\bibinfo{author}{\bibfnamefont{G.}~\bibnamefont{M\'{a}ndi}},
  \bibinfo{author}{\bibfnamefont{G.}~\bibnamefont{Teobaldi}}, \bibnamefont{and}
  \bibinfo{author}{\bibfnamefont{K.}~\bibnamefont{Palot\'{a}s}},
  \bibinfo{journal}{{P}rog. {S}urf. {S}ci.} \textbf{\bibinfo{volume}{90}},
  \bibinfo{pages}{223} (\bibinfo{year}{2015}).

\bibitem[{\citenamefont{Tersoff and Hamann}(1983)}]{Tersoff1983}
\bibinfo{author}{\bibfnamefont{J.}~\bibnamefont{Tersoff}} \bibnamefont{and}
  \bibinfo{author}{\bibfnamefont{D.}~\bibnamefont{Hamann}},
  \bibinfo{journal}{{P}hys. {R}ev. {L}ett.} \textbf{\bibinfo{volume}{50}},
  \bibinfo{pages}{1998} (\bibinfo{year}{1983}).

\bibitem[{\citenamefont{Seine et~al.}(1999)\citenamefont{Seine, Coratger,
  Carladous, Ajustron, Pechou, and Beauvillain}}]{Seine1999}
\bibinfo{author}{\bibfnamefont{G.}~\bibnamefont{Seine}},
  \bibinfo{author}{\bibfnamefont{R.}~\bibnamefont{Coratger}},
  \bibinfo{author}{\bibfnamefont{A.}~\bibnamefont{Carladous}},
  \bibinfo{author}{\bibfnamefont{F.}~\bibnamefont{Ajustron}},
  \bibinfo{author}{\bibfnamefont{R.}~\bibnamefont{Pechou}}, \bibnamefont{and}
  \bibinfo{author}{\bibfnamefont{J.}~\bibnamefont{Beauvillain}},
  \bibinfo{journal}{{P}hys. {R}ev. {B}} \textbf{\bibinfo{volume}{60}},
  \bibinfo{pages}{11045} (\bibinfo{year}{1999}).

\end{thebibliography}
\end{document}